\documentstyle[twocolumn,prb,aps,floats,epsfig]{revtex}
\begin{document}
\twocolumn[\hsize\textwidth\columnwidth\hsize\csname 
@twocolumnfalse\endcsname
\title{Piezoresistivity and conductance anisotropy
of tunneling-percolating systems} 
\author{C. Grimaldi$^1$, T. Maeder$^{1,2}$, P. Ryser$^1$, and S. Str\"assler$^{1,2}$} 
\address{$^1$ Institut de Production et Robotique, LPM,
Ecole Polytechnique F\'ed\'erale de Lausanne,
CH-1015 Lausanne, Switzerland}
\address{$^2$ Sensile Technologies SA, PSE, CH-1015 Lausanne, Switzerland}

\maketitle

\centerline \\

\begin{abstract}
Percolating networks based on interparticle tunneling conduction are 
shown to yield a logarithmic divergent piezoresistive response close
to the critical point as long as the electrical conductivity becomes
nonuniversal. At the same time, the piezoresistivity or, 
equivalently, the conductivity anisotropy 
exponent $\lambda$ remains universal
also when the conductive exponent is not, suggesting a purely geometric
origin of $\lambda$. We obtain these results by an exact solution of the 
piezoresistive problem on a Bethe lattice and by Monte Carlo calculations 
and finite-size scaling analysis on square lattices. 
We discuss our results in relation to the nature 
of transport for a variety of materials such as carbon-black--polymer composites and
RuO$_2$-glass systems which show nonuniversal transport properties and coexistence between
tunneling and percolating behaviors. 
\end{abstract}
\vskip 2pc ] 


\section{introduction}
\label{intro}

Tunneling-percolating network models have been suggested to properly
describe transport properties of disordered systems in which
tunneling coexists with percolation behaviors. Examples of such
systems are provided by various carbon-black--polymer composites, \cite{balb1}
for which, close to the metal-insulator transition, the conductivity $\sigma$ follows 
a percolation-like power law of the form: 
\begin{equation}
\label{sigma}
\sigma\simeq \sigma_0 (p-p_c)^t,
\end{equation}
where $\sigma_0$ is a prefactor, $p$ is 
the concentration of the 
conducting phase, $p_c$ is the percolation critical concentration, and $t$ is the
critical exponent.\cite{balb2} A clear indication that transport is
dominated by tunneling between carbon-black grains is provided by a large
strain and/or volume sensitivity of $\sigma$.\cite{sichel,heaney}
Similar situations are encountered also in organic conductors-polymer 
composites,\cite{zuppi} and thick-film resistors made of metal-oxide
conductive grains inhomogeneously embedded in a glassy matrix.\cite{carcia,tambo}
These latter compounds display quite large piezoresistive effects and
their use for sensor applications is widespread.\cite{prude1}

Modelling such systems in terms of tunneling-percolating networks resides on the
assumption that the exponential decay $\exp(-2 r/\xi)$ of the tunneling amplitude 
permits to retain only nearest neighbouring conducting grains. Since the
tunneling factor $\xi\propto 1/\sqrt{V}$ depends only on the potential barrier $V$
separating the grains, a modification
of the tunneling distance $r$ induced by an applied strain $\varepsilon$ 
can lead to a large variation of the total conductivity, and so to large values
of the piezoresistive coefficient $d\log(\sigma^{-1})/d\varepsilon$. For example both
carbon-black--polymer composites and RuO$_2$-based thick-film resistors
can display piezoresistive coefficients up to about $30$.\cite{sichel,prude1}

Although percolating network theories predict that the conductivity exponent
$t$ should be universal and equal to $t\simeq 2.0$ for 
three dimensional lattices,\cite{clerc}
tunneling-percolating systems display important deviations from 
universality. For example, values of $t$ up to about $t\simeq 6$ have been reported
for some kind of carbon-black composites,\cite{balb1,balb2}
while $t\simeq 4.0$ has been measured for RuO$_2$-based thick-film resistors.\cite{kusy}

Nonuniversality has been theoretically proposed to arise from specific 
conducting-insulating phases distributions such as in the Swiss-cheese model
where spherical insulating inclusions are introduced in a continuous 
conductor.\cite{halpe} An alternative explanation has been put forward specifically
for tunneling-percolating materials by arguing that, if the composite has a 
sufficiently wide distribution of tunneling distances, the distribution 
of bond conductances between conductive grains has a power-law divergence.\cite{balb3}
It is well known that such anomalous distribution can lead to nonuniversality.\cite{kogut}

In this paper we argue that the piezoresistive effect, {\it i. e.},
the sensitivity of $\sigma$ upon applied strain, could be a decisive tool 
to investigate the origin of nonuniversality.
We show that if nonuniversality of tunneling-percolating
systems is given by a diverging distribution of tunneling conductances,
then the piezoresistive response diverges logarithmically at the percolation
thresold, while at the same time the piezoresistivity anisotropy remains universal regardless
of the universality breakdown of the conductivity exponent $t$.
We obtain these results by an analytic solution of the piezoresistive problem
on a Bethe lattice and by Monte Carlo calculations and finite-size
scaling analysis on square lattices.

This paper is organized as follows. In the next section we introduce the lattice
model and the relevant quantities concerning the piezoresistive response. 
In section III we solve analytically the piezoresistive problem for a Bethe lattice
while in Section IV we investigate numerically the square lattice. A discussion of
our results in connection with real materials is presented in the last section where 
also the conclusions are drawn.

\section{the model}
\label{model}

Let us consider a random resistor network where the bond
conductivity  distribution is given by:
\begin{equation}
\label{eq1}
\rho(g)=p\, h(g)+(1-p)\delta (g),
\end{equation}
where $p$ is the fraction of bonds with finite conductivity $g$
with distribution $h(g)$. In the following, we implicitly assume that
the temperature is high enough to neglect other processes than intergrain
tunneling (grain charging effects, Coulomb repulsion). This is certainly a
good approximation for thick-film resistors which shows a very weak temperature
dependence of $\sigma$ already at room temperature.
Hence, within a tunneling-percolation framework,
$h(g)$ depends upon the distribution of tunneling distances between
nearest neighbouring grains. For narrow distributions, $h(g)$ is peaked
around $g=g_0\exp(-2a/\xi)$ where $a$ is the mean intergrain distance,
$\xi$ is the tunneling factor, and $g_0$ has
units of a conductivity. We assume that the main dependence upon the mean
intergain distance $a$ is all contained in the argument of the exponential and
for simplicity we set $g_0=1$.
In a first approximation, we can set 
\begin{equation}
\label{eq2}
h(g)=\delta (g-e^{-2a/\xi}),
\end{equation}
as an extreme case for narrow distribution for tunneling distances.
In this situation, close to the metal-insulator transition, transport
follows Eq.(\ref{sigma}) with universal exponent $t$ depending only 
upon dimensionality and all dependence upon tunneling distance $a$ is embodied
in the prefactor $\sigma_0$ which measures the average bond conductivity. 
Hence, we expect that a change of the tunneling distance $a$
induced by an applied external strain would affect only $\sigma_0$.

Let us consider now the case in which $N$ spherical particles of 
radius $R_0$ are placed at random so that the tunneling distance $r$ 
between two neighbouring grains fluctuates. When
the average distance $a$ between two adjacent grains is such that $a\gg R_0$,
Balberg argued that the salient feature of the resulting bond conductivity
distribution is captured by a power-law divergence at small $g$:\cite{balb3}
\begin{equation}
\label{eq3}
h(g)=(1-\alpha)g^{-\alpha},
\end{equation}
where $\alpha=1-\xi/2a$ and the
prefactor $1-\alpha$ assures the correct normalization of $h(g)$ and
$\rho(g)$. As first shown by Kogut and Straley,\cite{kogut} for values of $\alpha$ 
larger than some dimensionality dependent critical value $\alpha_c$, universality
breaks down and the transport exponent $t$ becomes $\alpha$-dependent:
$t\rightarrow t(\alpha)>t$. The important point here is that since $\alpha$
depends on the mean intergrain distance $a$, then the exponent $t(\alpha)$ can be affected
by an applied external strain $\varepsilon$ and this would lead to an anomalous 
piezoresistive response.

To investigate this issue, let us assume that the network
is embedded in a homogeneous elastic medium and that the elastic
coefficients of the network and the medium are equal. Moreover we set
the bond directions parallel to the axes of a
$D$-dimensional cubic lattice. In this situation, if we apply an
uniaxial strain $\varepsilon$ along, for example, the $x$-axis, then the
tunneling distance for a bond directed along $x$ changes to 
$a\rightarrow a_x=a(1+\varepsilon)$, while the bonds directed along the
other orthogonal axes remain unchanged: $a_i=a$ for $i\neq x$.
The strain-induced change of the tunneling distances leads therefore
to anisotropic bond conductivity
distributions $\rho_x(g)$ and $\rho_{i\neq x}(g)=\rho(g)$. 
For an external electric field $E_i$
applied along the $i$-axis and up to linear order in $\varepsilon$, 
the resulting network conductivities $\sigma_i$ 
are $\sigma_x=\sigma(1-\Gamma_\parallel\varepsilon)$ and
$\sigma_{i\neq x}=\sigma(1-\Gamma_\perp\varepsilon)$ where $\sigma$ is the unstrained
conductivity and
\begin{equation}
\label{eq4}
\Gamma_\parallel=\frac{d\ln\sigma_x^{-1}}{d\varepsilon}; \,\,\,\,\,
\Gamma_\perp=\frac{d\ln\sigma_i^{-1}}{d\varepsilon}\,\,\,(i\neq x),
\end{equation}
are the longitudinal and transverse piezoresistive coefficients, respectively.
These two quantities completely define the piezoresistive properties of the
network. For example, the isotropic (or hydrostatic) piezoresistive response $\Gamma$
obtained by applying equal strain $\varepsilon$ along all bonds directions is:\cite{grima1}
\begin{equation}
\label{hydro}
\Gamma=\frac{d\ln\sigma^{-1}}{d\varepsilon}=\Gamma_\parallel+(Z/2-1)\Gamma_\perp,
\end{equation}
where $Z$ is the coordination number, 
while informations about the tortuosity the current has in flowing through
the network are given by the piezoresistive anisotropy factor defined as:
\begin{equation}
\label{eq10}
\chi=\frac{\Gamma_\parallel-\Gamma_\perp}{\Gamma_\parallel},
\end{equation}
which measures the degree of macroscopic transport anisotropy.\cite{grima1}
Close to the percolation thresold, $\chi$ displays a power-law of the form:\cite{grima1}
\begin{equation}
\label{chipower}
\chi\sim (p-p_c)^\lambda,
\end{equation}
where the exponent 
$\lambda$ is the same of that governing criticality of
the conductivity anisotropy $A=1-\sigma_y/\sigma_x$ of random resistor
networks with anisotropic bond conductances.\cite{shklo,sary,sahimi} 
In the present case, in fact,
bond anisotropy is induced by the applied uniaxial strain and, since $\lambda$
is independent of the degree of bond anisotropy,\cite{lobb} $A$ and $\chi$ 
have the same critical behavior with the same exponent $\lambda$.

Having introduced the main quantities defining the piezoresistive properties of
random resistor networks, let us now discuss qualitatively the effects of
universality breakdown induced by the diverging bond conductances distribution function
Eq.(\ref{eq3}). The anisotropy factor $\chi$ and its exponent $\lambda$ cannot
be accounted in a simple way without explicitly solving the anisotropy problem,
which is done in the following sections. However, the hydrostatic piezoresistive
response $\Gamma$ can be simply evaluated by noticing that, by definition, it
is obtained by changing the tunneling distance $a$ to $a(1+\varepsilon)$ for all
bond directions. Hence $\Gamma$
can be readily found by differentiating Eq.(\ref{sigma}) with respect to $\varepsilon$:
\begin{equation}
\Gamma=
\left\{\begin{array}{lcc}
\Gamma_0 & \hspace{5mm} & \alpha<\alpha_c \\
\Gamma_0+(1-\alpha)\ln\left(\frac{1}{p-p_c}\right)t'(\alpha) & 
\hspace{5mm} & \alpha\ge\alpha_c
\end{array}
\right.,
\label{hydro2}
\end{equation}
where $\Gamma_0=-d\ln\sigma_0^{-1}/d\varepsilon$. For $\alpha<\alpha_c$ transport
is universal and the piezoresistive response is governed solely by the strain-dependence
of $\sigma_0$ in Eq.(\ref{sigma}) leading to the $p$-independent factor $\Gamma_0$.
Of course, distribution functions for occupied bonds like Eq.(\ref{eq2}) would
trivially lead to the same qualitative result, {\it i.e.}, a piezoresistive
response independent of the concentration of occupied bonds.
Instead, as soon as transport
becomes nonuniversal ($\alpha\ge\alpha_c$), the tunneling-percolation model 
of Balberg Eq.(\ref{eq3}) predicts
a logarithmic divergence of the hydrostatic piezoresistive response.
Note that such an anomalous behavior is expected also for distributions more
complicated than Eq.(\ref{eq3}) as long as their asymptotic behavior for $g\rightarrow 0$
has a power-law divergence with exponent depending upon the tunneling distance.

In the next sections we shall verify the correctness of Eq.(\ref{hydro2}) and calculate the
$p$-dependence of $\chi$ by considering two quite dinstinct cases: the Bethe lattice model,
which is paradigmatic of high-dimensionality lattices for which transport is 
governed by mean-field exponents,\cite{stauffer} and the two-dimensional square lattice.

\section{Bethe lattice}
\label{bethe}

The conductivity problem on a Bethe lattice, or Cayley tree, model 
has been considered and solved in Refs.[\onlinecite{stinch,straley1}]. As we show below,
for this model the concentration dependence of the piezoresistive response 
in the critical regime can be obtained
analytically for both distributions Eqs.(\ref{eq2},\ref{eq3}).
For simplicity in the following we shall consider a Bethe lattice with
coordination number $Z=4$ ($p_c=1/(Z-1)=1/3)$). According to 
Ref.[\onlinecite{straley2}], the distribution $P_i(\sigma)$ of conductivities
from an arbitrary bond directed along $i=x,y$ to infinity and the
current distributions $J_i(\sigma)$ induced by an applied electric field $E_i$ satisfy
the following coupled non-linear integral equations:
\begin{eqnarray}
\label{eq5}
P_x(\sigma_1)=&&\int_0^1dg\rho_x(g)\int d\sigma_2 d\sigma_3 d\sigma_4
P_x(\sigma_2)P_y(\sigma_3)P_y(\sigma_4) \nonumber \\
&&\times\delta\!\left[\sigma_1-\frac{g(\sigma_2+\sigma_3+\sigma_4)}
{g+\sigma_2+\sigma_3+\sigma_4}\right],
\end{eqnarray}
\begin{eqnarray}
\label{eq6}
J_x(\sigma_1)&&=E_x\sigma_1 P_x(\sigma_1)+\int_0^1dg\rho_x(g)
\int d\sigma_2 d\sigma_3 d\sigma_4 \nonumber \\
&&\frac{g\,J_x(\sigma_2)P_y(\sigma_3)P_y(\sigma_4)}{g+\sigma_2+\sigma_3+\sigma_4}
\delta\!\left[\sigma_1-\frac{g(\sigma_2+\sigma_3+\sigma_4)}
{g+\sigma_2+\sigma_3+\sigma_4}\right]. 
\end{eqnarray}
The corresponding equations for $P_y(\sigma)$ and $J_y(\sigma)$ are obtained
from Eqs.(\ref{eq5},\ref{eq6}) by substituting $x$ with $y$ (and vicerversa), 
and the total conductivities $\sigma_i$ are obtained from:
\begin{equation}
\label{eq7}
\sigma_x=-2\int_0^\infty dz J_x(z)P_y(z)\frac{d[P_x(z)P_y(z)]}{dz},
\end{equation}
where $P_i(z)$ and $J_i(z)$ are the Laplace transforms of $P_i(\sigma)$
and $J_i(\sigma)$, respectively.\cite{straley2}

Let us consider first the case of a narrow tunneling-distances distribution,
Eqs.(\ref{eq1},\ref{eq2}), for which, in the absence of applied strain,
the conductivity is $\sigma=6\exp(-2a/\xi) (p-p_c)^t$ for $|p-p_c|\ll 1$ where
$t=3$ is the universal transport exponent.\cite{straley1}
The effect of an applied uniaxial strain, $\varepsilon\neq 0$, can be readily
found by following
Ref.[\onlinecite{straley2}]. In the critical region, we find that
the piezoresistive coefficients Eq.(\ref{eq4}) reduce to:
\begin{equation}
\label{eq8}
\Gamma_{\parallel (\perp)}=\frac{a}{\xi}\left[1+(-)\frac{15}{16}(p-p_c)\right]. 
\end{equation}
The above result captures the essential physics at the basis of the
piezoresistive response of percolating networks.\cite{grima1} For $p>p_c$, 
$\Gamma_\parallel>\Gamma_\perp$ because the strain-sensitivity is stronger
for sample conductivities measured along the direction of the applied strain.
As $p$ moves towards $p_c$, 
the current currying paths become more and more tortuous decreasing therefore
the macroscopic anisotropy induced by $\varepsilon$ until at $p=p_c$
the longitudinal and transverse piezoresistive coefficients become equal.\cite{grima1}
From Eq.(\ref{eq8}), we find that the piezoresistive anisotropy $\chi$, Eq.(\ref{eq10}), 
goes to zero as:
\begin{equation}
\label{eq10bis}
\chi=\frac{15}{8}(p-p_c).
\end{equation}
Hence the anisotropy exponent for the Bethe lattice is $\lambda=1$ in accord
with previous results.\cite{straley2}

Now, we consider how the piezoresistive response of a Bethe lattice changes 
when the bond conductances have a power-law distribution as in Eqs.(\ref{eq1},\ref{eq3}).
In doing so, we generalize the procedure described in Ref.\onlinecite{kogut}
to the anisotropic bond conductance case of Eqs.(\ref{eq5}-\ref{eq7})
and perform an expansion in powers of $\varepsilon$. To obtain the
piezoresistive coefficients, it is sufficient to keep only terms up to linear order in
$\varepsilon$. Here we stress that the resulting piezoresistive response for
the Bethe lattice model depends
crucially on the sign of $\alpha$. For $\alpha <0$, corresponding to the quite
unphysical relation $a<\xi/2$, conductivity is universal with $t=3$,\cite{kogut} and the
piezoresistive response can be easily shown to behave qualitatively as in the
binary distribution case discussed above (finite values of $\Gamma_{\parallel (\perp)}$
at $p=p_c$ and $\lambda=1$). Instead when $\alpha>0$, the piezoresistive response
changes qualitatively as discussed in the following.

From Eq.(\ref{eq5}) we find that for $\alpha>0$ the $p-p_c$ dependence 
of the Laplace transforms of $P_x(\sigma)$ and $P_y(\sigma)$ is:
\begin{equation}
\label{Pxy}
P_{x(y)}(z)=1-\delta\,f(\omega)+\delta\ln\delta
[g(\omega)+(-)\delta\,g_1(\omega)]\varepsilon,
\end{equation}
where $\omega=\delta^\frac{1}{1-\alpha}z$ and $\delta=(p-p_c)/p_c$. The
functions $f(y)$, $g(y)$ and $g_1(y)$ satisfy coupled integral equations,
but their explicit expressions are not of interest here.
By applying the same procedure to Eq.(\ref{eq6}) we find:
\begin{equation}
\label{Jxy}
J_{x(y)}(z)=\frac{3}{2}\delta^{1+\frac{1}{1-\alpha}}E_x\{
f'(\omega)-\ln\delta[g'(\omega)+(-)\delta\,g_2(\omega)]\varepsilon\}.
\end{equation} 
Finally, by substituting these results into Eq.(\ref{eq7}) we obtain
\begin{equation}
\label{eq15}
\sigma_{x(y)}=\delta^{3+\frac{\alpha}{1-\alpha}}
\{a_0-a_1\varepsilon- \ln (\delta^{-1})[a_2+(-)a_3\delta] \varepsilon\}, 
\end{equation}
where $a_0,\ldots, a_3$ are positive functions of $\alpha$. For $\varepsilon=0$,
we obtain the result of Kogut and Straley:\cite{kogut} $\sigma_{x(y)}=\sigma\sim (p-p_c)^{t(\alpha)}$
with $t(\alpha)=3+\alpha/(1-\alpha)$, indicating that transport is non-universal
for $\alpha>\alpha_c=0$. In this regime, the piezoresistive coefficients are obtained
from the terms of Eq.(\ref{eq15}) proportional to $\varepsilon$:
\begin{equation}
\label{eq17}
\Gamma_{{\parallel}(\perp)}=\frac{a_1+\ln (\delta^{-1})[a_2+(-) a_3\delta]}{a_0}
\sim \ln \! \left( \frac{1}{p-p_c} \right).
\end{equation}
We have arrived therefore at the result that as long as
the tunneling-distance distribution is such that transport becomes non-universal,
then the piezoresistive response diverges logarithmically as $p\rightarrow p_c$.
This must be contrasted with the finite value of $\Gamma_\parallel$ and
$\Gamma_\perp$ at $p=p_c$, Eq.(\ref{eq8}), obtained from the simple binary 
distribution Eqs.(\ref{eq1},\ref{eq2}). From Eq.(\ref{eq17}) we obtain also that the 
piezoresistive anisotropy $\chi$, Eq.(\ref{eq10}), reduces to
\begin{equation}
\label{chi2}
\chi=\frac{2a_3}{a_2}\delta\propto (p-p_c). 
\end{equation}
Hence $\chi$ goes to zero with universal exponent $\lambda=1$
irrespectively of the logarithmic divergence of $\Gamma_{\parallel (\perp)}$
and of the universality breakdown of $\sigma$. 

\begin{figure}[t]
\protect
\centerline{\epsfig{figure=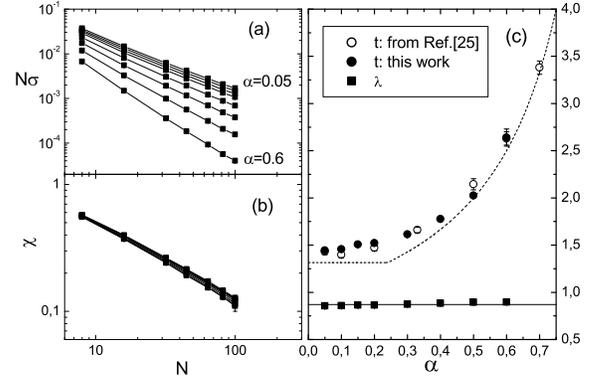,width=20pc,clip= }}
\caption{Unstrained conductivity $\sigma$ (a) and piezoresistive anisotropy 
$\chi$ (b) calculated at $p=p_c=1/2$ as function of the
width $N$ of the strip for $\alpha=0.05,0.1,0.15,0.2,0.3,\ldots 0.6$. (c): solid circles: conductivity exponent $t$ as a function of 
the distribution exponent $\alpha$. Empty circles: results from Ref.\protect\onlinecite{octavio}.
Dashed curve is the theoretical estimate $t(\alpha)=1.3158$ for $\alpha<\alpha_c=0.24$
and $t(\alpha)=1/(1-\alpha)$ for $\alpha>\alpha_c$.
Solid squares: conductivity (or piezoresistivity) anisotropy exponent $\lambda$.
Solid line: average value $\lambda=0.87$.}
\label{fig1}
\end{figure}

\begin{figure}[t]
\protect
\centerline{\epsfig{figure=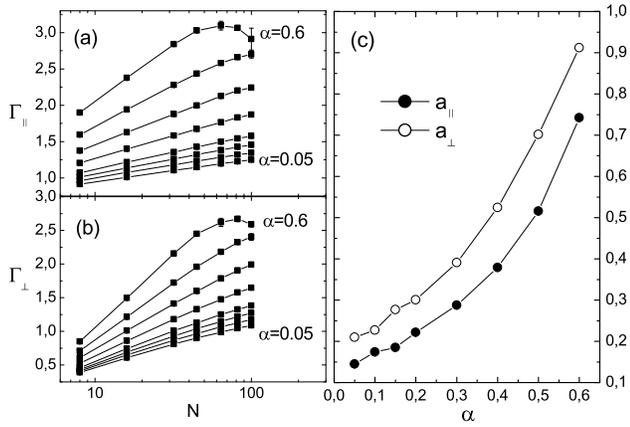,width=20pc,clip= }}
\caption{(a) and (b): longitudinal and transverse piezoresistive coefficients
as a function of $N$ for different values of the power-law distribution 
exponent $\alpha$. (c): prefactor of $\Gamma_{\parallel (\perp)}=
a_{\parallel (\perp)}\ln (N)$ as a function of $\alpha$.}
\label{fig2}
\end{figure}

\section{square lattice}
\label{square}

The results of the last section have been derived for a Bethe lattice
and are therefore in general relevant for high-dimensionality ($D\ge 6$)
lattices.\cite{stauffer}
To investigate the piezoresistive response for low dimensional networks,
we consider now a square lattice. 
We employ the transfer-matrix method of Derrida and
Vannimenus which permits to calculate exactly the conductivity of
a long strip of width $N$ (along $x$) and length $L\gg N$ (along $y$).\cite{derrida} 
The distribution of the $N\times L$
conductors follows Eq.(\ref{eq1},\ref{eq3}) 
and the longitudinal and transverse 
piezoresistive coefficients are obtained by setting $a_x=a(1+\varepsilon)$, $a_y=a$
and $a_x=a$, $a_y=a(1+\varepsilon)$, respectively, with $\varepsilon=0.001$.
For each value of $\alpha >0$ at the percolation thresold $p=p_c=1/2$, we perform 
the calculations at
$N=8\,(L=1\times 10^8)$,  $N=16\,(L=7.5\times 10^7)$, $N=32\,(L=2.5\times 10^7)$,
$N=45\,(L=2\times 10^7)$,  $N=64\,(L=1\times 10^7)$, $N=82\,(L=1\times 10^7)$,
and $N=100\,(L=0.8\times 10^7)$.

In Fig.\ref{fig1} and Fig.\ref{fig2} we show our results for the unstrained conductivity 
$\sigma$, the longitudinal and transverse piezoresistive coefficients 
$\Gamma_{\parallel (\perp)}$, and the piezoresistive anisotropy $\chi$.
From the $\sigma$-{\it vs}-$N$ data of Fig.\ref{fig1}(a) we extract the
conductivity exponent $t$ from the finite-size scaling relation
$\sigma(N)=c_1 N^{-t/\nu}(1+c_2/N)$ where $\nu=\frac{4}{3}$ is the correlation
length exponent.\cite{zabo} The resulting $t$-values (solid circles) are reported in Fig.\ref{fig1}(c) 
as function of the power-law distribution exponent $\alpha$ 
of Eq.(\ref{eq3}).
For comparison, we report also the $t$ values obtained in 
Ref.\onlinecite{octavio} by a different numerical method (empty circles). As expected,
for $\alpha$ sufficiently large, the $t$-exponent increases well beyond its
universal value $t\simeq 1.3$,\cite{zabo,octavio} signalling breakdown of universality. 
According to
Refs.\onlinecite{octavio,machta}, universality is lost for $\alpha>\alpha_c=0.24\pm 0.08$,
and the conductivity exponent should follow the relation $t(\alpha)=1/(1-\alpha)$
reported in Fig.\ref{fig1}(c) by the dashed curve. Our numerical results, as those of 
Ref.\onlinecite{octavio}, agree well with the theoretical expectations for large values of
$\alpha$ but somehow overestimates the $t$-values for $\alpha\sim \alpha_c$. 
This deviation from the theoretical expectations should be due to
finite-size effects amplified by the presence of two competing 
fixed points.\cite{octavio,machta}

In contrast to the large $\alpha$-dependence of $\sigma(N)$, the data for different
$\alpha$ values of the
piezoresistive anisotropy $\chi$ reported in Fig.\ref{fig1}(b) collapse all
in a single curve. Hence, the piezoresistivity (or conductivity)
anisotropy exponent $\lambda$ [filled squares in Fig.\ref{fig1}(c)]
estimated from the finite-size scaling relation
$\chi(N)=c_3 N^{-\lambda/\nu}(1+c_4/N)$ does not show appreciable variations 
over the entire range of $\alpha$-values considered.\cite{note}
This is in striking contrast with $\lambda=t-\beta_B$, where $\beta_B\simeq 0.48$
is the fraction of conducting bones in the backbone, 
conjectured to hold true for $D=2$ in Ref.\onlinecite{sahimi}.
As for the Bethe lattice case, the universality of $\lambda$ also for $\alpha>\alpha_c$
suggests that this exponent depends only upon the geometry of the conducting
cluster and is not influenced by $t$.

The results for the piezoresistive coefficients $\Gamma_\parallel$ and $\Gamma_\perp$ as 
functions of $N$ are reported in Fig.\ref{fig2}(a) and Fig.\ref{fig2}(b), respectively.
With the exclusion of the large $N$ values for $\alpha=0.6$, which we think are affected by
too small values of $L$,
$\Gamma_\parallel$ and $\Gamma_\perp$ follow approximatively a $\ln (N)$ behavior.
This signals a $\ln [(p-p_c)^{-1}]$ divergence of the piezoresistive coefficients 
as $p\rightarrow p_c$. How the logarithmic divergence depends upon $\alpha$
is studied in Fig.\ref{fig2}(c) where we plot the prefactors $a_\parallel$ and
$a_\perp$ of the finite-size scaling law: $\Gamma_{\parallel (\perp)}=
a_{\parallel (\perp)}\ln (N) (1+\cdots)$. The correction-to-scaling terms $\cdots$
which best fitted the data where proportional to $1/N$ and to $1/\ln (N)$.
Clearly, $a_{\parallel (\perp)}$ is a monotonous
increasing function of $\alpha$ indicating that the logarithmic divergence of
$\Gamma_{\parallel (\perp)}$ is stronger for $\alpha$ larger. 
In analogy with the results on the Bethe lattice, we would expect
that $a_{\parallel (\perp)}$ vanishes for $\alpha <\alpha_c$, so that 
$\Gamma_{\parallel (\perp)}$ has a finite limit at $p=p_c$. However the
$a_{\parallel (\perp)}$ data of Fig.\ref{fig2}(c) are small but nonzero
even for $\alpha <\alpha_c =0.24$. We think that this is due to the same finite-size
errors affecting the unstrained conductivity exponent, Fig.\ref{fig1}(c), which lead
to a spurious $\alpha$-dependence of $\sigma$ and consequently to 
$a_{\parallel (\perp)} \neq 0$.

\section{discussion and conclusions}
\label{concl}

Let us discuss now the applicability of our theory to real materials.
The thermal expansion effect on the resistance of nonuniversal ($t\simeq 3.0$)
carbon-black--polymer composite reported in Ref.\onlinecite{heaney} provides
a first indirect clue. From the 
resistance-{\it vs}-volume data we have extracted a hydrostatic 
piezoresistive coefficient $\Gamma$ of about
$30$ which keeps increasing as the volume is expanded by the temperature. If volume
expansion effectively reduces the carbon-black concentration, we obtain a piezoresistive
coefficient enhancement as the concentration $p$ moves down to its critical value.
However, due to the uncontrolled effect of polymer melting on the microstructure of
the composite, we have been unable to single out any logarithmic divergence of the
piezoresistivity response. 

\begin{figure}[t]
\protect
\centerline{\epsfig{figure=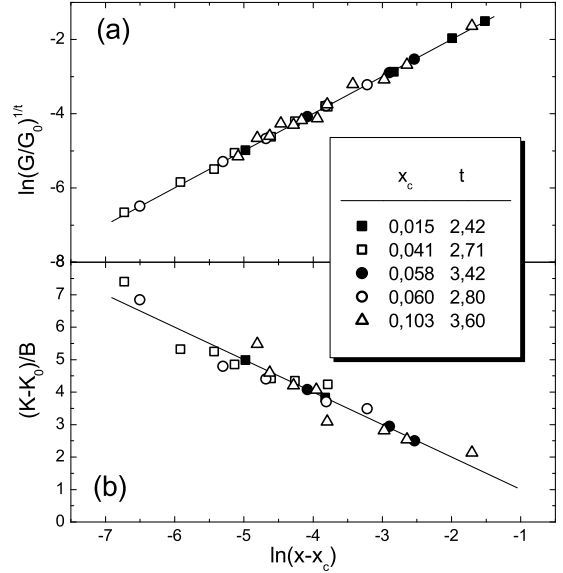,width=20pc,clip= }}
\caption{Conductance $G$ (a) and piezoresistance $K$ (b) of RuO$_2$-glass composites
for various RuO$_2$ volume concentrations $x$. 
Open symbols: Ref.\protect\onlinecite{carcia}. 
Solid symbols: Ref.\protect\onlinecite{tambo}.}
\label{fig3}
\end{figure}

A much clearer situation is found for RuO$_2$-based thick-film resistors.
In Fig.\ref{fig3}(a) we report conductance $G$ measurements on several
RuO$_2$-glass composites for different values of the metal volume 
concentration $x$.\cite{carcia,tambo}
When $\ln(G/G_0)^{1/t}$ is plotted as a function of $\ln(x-x_c)$, where $G_0$
is a prefactor and $x_c$ is the critical volume concentration, the whole set of
data collapses into a single straight line indicating a power law of the form:
\begin{equation}
\label{exp1}
G\simeq G_0 (x-x_c)^t.
\end{equation}
In Fig.\ref{fig3}(a), the different symbols refer to different relative particles
sizes of the glass and RuO$_2$ and
the values of $x_c$ and $t$ which best fit the experimental data are reported in the 
inset. Clearly, all composites display values of $t$ much higher than the
three dimensional universal value $t\simeq 2.0$.\cite{clerc}

The origin of such universality breakdown is investigated in Fig.\ref{fig3}(b)
where we plot the $x$-dependence of the longitudinal piezoresistance coefficient 
$K=d\ln G^{-1}/d\varepsilon$ obtained by cantilever bar 
measurements.\cite{carcia,tambo} 
The measured $K$ values fit reasonably well with a logarithmically divergent
function of the form:
\begin{equation}
\label{exp2}
K\simeq K_0+B\ln\left(\frac{1}{x-x_c}\right),
\end{equation}
represented in the $(K-K_0)/B$-{\it vs}-$\ln(x-x_c)$ plot of
Fig.\ref{fig3}(b) by the straight line. 
Since all the samples lie well within the critical
region [Fig.\ref{fig3}(a)], we expect that the piezoresistive anisotropy
$\chi$ is sufficiently small to regard $K$ as a good approximation of the
isotropic (hydrostatic) piezoresistance coefficient.\cite{note2} Hence,
the data of Fig.\ref{fig3}
are fully consistent with Eq.(\ref{hydro2}) for $\alpha\ge \alpha_c$ and provide
a rather good example of the effect we have described in this paper.

Unfortunately we are not aware of reported measurements recording 
$\chi$ as a function of $x$, so that the universality of $\lambda$ claimed here
even for $\alpha\ge\alpha_c$ cannot be verified. 
However the measurements reported in of Ref.\onlinecite{hrovat} show 
that $\chi$ decreases as the sheet resistance of commercial RuO$_2$-based thick-film resistors 
increases. This is in qualitative accord with $\chi\sim (x-x_c)^\lambda$ if higher
resistance values are due to lower RuO$_2$ concentrations.

In summary, we have shown by means of analytical and numerical results that 
when the tunneling exponent $\alpha=1-\xi/2a$ of the power-law 
distribution Eq.(\ref{eq3}) is such that transport becomes non-universal, the piezoresistive
response changes drastically leading to a logarithmic divergence of the piezoresistive
coefficients as $p\rightarrow p_c$. In addition, we have demonstrated that despite 
of the universality breakdown of transport,
the conductivity anisotropy exponent $\lambda$ remains universal. 
These features seem to be quite robust and calls for
experimental verifications on systems like carbon-black--polymer composites and
thick-film resistors for which tunneling-percolation mechanism of transport have been
proposed and non-universality has been reported.\cite{balb1,balb2,balb3,kogut}.
Earlier experimental results on carbon-black--polymer composites,\cite{heaney} and especially
RuO$_2$ thick-film resistors,\cite{carcia,tambo} seem to indicate that indeed these systems
are in the diverging tunneling conductance distribution regime.


\begin{references}
\bibitem{note}
The average value of $\lambda$ over the different values of $\alpha$
is $\lambda=0.87\pm 0.02$ which is only
somewhat larger than $\lambda=0.819\pm 0.011$ estimated in Ref.\cite{sahimi}.
\bibitem{note2}
The piezoresistance ($K$) and the
piezoresistive ($\Gamma$) coefficients differ by an additive constant which 
depends only upon the geometry of the sample.\cite{grima1}
\end{references}
\end{document}